%Paper: hep-th/9311133
%From: Cenalo Vaz <cvaz@mozart.si.ualg.pt>
%Date: Mon, 22 Nov 1993 19:23:45 GMT

%Hardcopies of figures available on request.
\input phyzzx
\vsize 9.25in
\hsize 6.3in
\def\vol{\int d^2 x {\sqrt{-g}}}
\def\scrrp{{\cal I}_R^+}
\def\scrlp{{\cal I}_L^+}

\rightline{UATP-93/04}
\rightline{October 1993}
\vskip 0.2in
\centerline{\seventeenbf Formation and Evaporation of a Naked
Singularity}
\centerline{\seventeenbf in 2 d Gravity}
\vskip 0.75in
\centerline{\caps Cenalo Vaz}
\centerline{\it Unidade de Ci\^encias Exactas e Humanas}
\centerline{\it Universidade do Algarve}
\centerline{\it Campus de Gambelas, P-8000 Faro, Portugal}
\vskip 0.2in
\centerline{and}
\vskip 0.2in
\centerline{\caps Louis Witten}
\centerline{\it Department of Physics}
\centerline{\it University of Cincinnati}
\centerline{\it Cincinnati, OH 45221-0011, U.S.A.}
\vskip 0.75in
\centerline{\bf \caps Abstract}
\vskip 0.2in

We describe a classical configuration of conformal matter forming
a naked singularity and discuss its subsequent Hawking evaporation
within the context of two dimensional dilaton gravity.  The one
loop analysis is credible for a large mass naked singularity and
suggests the existence of a weak cosmological censorship that would
cause it to explode into radiation upon forming.
\vfill
\eject

The discrepancy between its classical and quantum histories seems
to require a fundamental revision of our understanding of black
hole physics.  Perhaps more importantly, it raises serious
questions of consistency and interpretation in quantum gravity.
Hawking's${}^1$ pioneering discovery that a black hole is not in
fact quite ``black'' but, via quantum effects actually radiates
away its energy in the form of thermal radiation at a
characteristic temperature, has inspired a great amount of
work${}^2$ attempting both to gain some insight into the ultimate
fate of these objects and to understand the dilemma introduced by
the apparent loss of quantum coherence${}^3$ in the process.  Most
attempts have focused on toy models of two dimensional
gravity,${}^4$ some inspired by string theory,${}^{5,6}$ in the
hope that some of these issues can be resolved while the theory yet
retains some of the important features of its four dimensional
(Einstein) counterpart.  While black holes have received a great
deal of attention, the formation and fate of naked singularities
remains comparatively neglected and obscure.  It turns out that
some models of two dimensional gravity provide simplified arenas in
which to study naked singularities.  In this paper we investigate
a modification, admitting positive energy naked singularities, of
a recently proposed model${}^{7,8}$ for a full quantum treatment of
two dimensional black hole evaporation.  Our aim is to describe
qualitatively the history of a naked singularity within the context
of this model.  This represents therefore a preliminary step to a
detailed quantum treatment of naked singularities, at least in two
dimensions.

The model we consider is described by the action
$$S~~ =~~ \vol \left[ e^{-2\phi} \left( -~ R~ +~ 4 (\nabla \phi)^2~
+~ \Lambda \right)~ -~ {1 \over 2}~ \sum_i (\nabla f_i)^2
\right]\eqno(1)$$
where $\Lambda$ is the cosmological constant, the $f_i(x)$ are $N$
conformally coupled matter fields and $\phi$ is the dilaton. The
model with positive $\Lambda$ was considered in ref (7) and shown
to be unstable against gravitational collapse, admitting black hole
solutions produced by incoming $f-$ waves.  However, quantum
effects in the one loop approximation made the $f-$ wave radiate
away all its energy even before the formation of a horizon, so that
black holes do not appear in the quantum spectrum.  As we show, the
model with negative $\Lambda$ admits naked singularities, and their
quantum evolution is very different.  They are formed, but do not
survive for long, radiating away their energy in a burst.  The
action in (1), with $\Lambda > 0$, was shown to arise as the
effective action describing the radial modes of extremal dilatonic
black holes in four dimensions and, without the $f-$ fields, from
string theory to the lowest order in world sheet perturbation
theory.  Here it will be considered simply as a model of two
dimensional gravity, and we will take $\Lambda = - 4 \lambda^2$
henceforth.  Our sign conventions throughout are those of
Weinberg${}^{9}$.

Varying (1) with respect to the metric $g^{\mu\nu}$ gives the
gravitational equation of motion
$$\eqalign{0~~ =~~ {\cal T}_{\mu\nu}~~ &=~~ e^{-2\phi}~ \left[ 2
\nabla_\mu \nabla_\nu \phi~ -~ {1 \over 2} e^{2\phi} \sum_i
\nabla_\mu f_i \nabla_\nu f_i \right. \cr &~~~~~~~~ \left. +~~
g_{\mu\nu}~ \left( -2 \nabla^2 \phi~ +~ 2 (\nabla \phi)^2~ +~
2\lambda^2~ +~ {1 \over 4} e^{2\phi} \sum_ i (\nabla f_i)^2 \right)
\right] \cr} \eqno(2)$$
where $\nabla_\mu$ is the covariant derivative with respect to
$g_{\mu\nu}$.  The three equations above form a set of constraints
on the allowable solutions of the field equations.  Passing to the
conformal gauge, in which the metric has the form
$$g_{\mu\nu}~~ =~~ e^{2\rho} \eta_{\mu\nu}, \eqno(3)$$
(1) may be re-expressed as
$$S~~ =~~ \int d^2x \left[ e^{-2\phi} \left( - 2\nabla^2 \rho~ +~
4 (\nabla \phi)^2~ -~ 4 \lambda^2 e^{2\rho} \right)~ -~ {1 \over 2}
\sum_i (\nabla f_i)^2 \right], \eqno(4)$$
where all derivatives are with respect to the flat metric
$\eta_{\mu\nu}$ and it is understood that the constraints in (2)
are satisfied fields above. One then obtains the following
equations of motion
$$\eqalign{2 \nabla^2 \phi~ -~ \nabla^2 \rho~ -~ 2 (\nabla \phi)^2~
-~ 2 \lambda^2 e^{2\rho}~~ &=~~ 0 \cr \nabla^2 \phi~ -~ 2 (\nabla
\phi)^2 ~  -~ 2 \lambda^2 e^{2\rho}~~ &=~~ 0 \cr \nabla^2 f_i~~
&=~~ 0. \cr} \eqno(5)$$
The solutions of interest are most readily obtained in light-cone
coordinates $x^\pm = x^0 \pm x^1$ which we use hereafter.  In the
conformal gauge the constraint equations reduce to
$$\eqalign{{\cal T}_{++}~~ &=~~ e^{-2\phi} \left[ 2 \partial_+
\partial_+ \phi~ -~ 4 \partial_+ \phi \partial_+ \rho \right]~ -~
{1 \over 2}~ \sum_i \partial_+ f_i \partial_+ f_i~~ =~~ 0 \cr {\cal
T}_{--}~~ &=~~ e^{-2\phi} \left[ 2 \partial_- \partial_-\phi~ -~ 4
\partial_- \phi \partial_- \rho \right]~ -~ {1 \over 2}~ \sum_i
\partial_- f_i \partial_- f_i~~ =~~ 0 \cr {\cal T}_{+-}~~ &=~~ e^{-
2\phi} \left[ -2 \partial_+ \partial_- \phi~ +~ 4 \partial_+ \phi
\partial_- \phi~ -~ \lambda^2 e^{2\rho} \right]~~ =~~ 0
\cr}\eqno(6)$$
and the equations of motion are
$$\eqalign{& -~ 4\partial_+ \partial_- \phi~ +~ 4 \partial_+ \phi
\partial_- \phi~ +~ 2 \partial_+ \partial_- \rho~ -~ \lambda^2
e^{2\rho}~~ =~~ 0 \cr & -~ 2 \partial_+ \partial_- \phi~ +~ 4
\partial_+ \phi \partial_- \phi~ -~ \lambda^2 e^{2\rho}~~ =~~ 0 \cr
& \partial_+ \partial_- f_i~~ =~~ 0. \cr} \eqno(7)$$
The first two equations in (7) imply that the $\rho (x)$ is the
same as $\phi(x)$ up to a harmonic function, $h(x)$.  A choice of
$h(x)$ amounts to a choice of coordinate system, because the choice
of conformal gauge does not fix the conformal subgroup of
diffeomorphisms.  Fixing this gauge freedom by taking $h(x) = 0$
the general solution satisfying the constraints has the form ($e^{-
2\phi}~ =~ e^{-2\rho}~ =~ \sigma$)
$$\sigma~~ =~~ \lambda^2 x^+ x^-~~ +~~ {M \over \lambda} \eqno(8)$$
where $M$ is a positive constant which, as we show below, is the
Bondi${}^{12}$ energy of the singularity.

When $M = 0$, the metric is flat and the dilaton is linear in the
spatial coordinate.  This is the linear dilaton vacuum and has
appeared before in various studies.  When $M \neq 0$, the metric is
$g_{\mu\nu} = e^{2\phi} \eta_{\mu\nu}$, and the curvature scalar
$$\eqalign{R~~ &=~~ +~ 2 e^{-3\rho} \nabla^2 e^\rho~~ -~~ 2 e^{-
4\rho} (\nabla e^\rho)^2 \cr &=~~ +~ 4~ \left[ \partial_+
\partial_-\sigma ~~ -~~ {{\partial_+ \sigma \partial_- \sigma}
\over \sigma} \right] \cr &=~~ +~ {{4 \lambda M} \over \sigma},
\cr} \eqno(9)$$
is singular at $\sigma(x) = 0$.  The Penrose diagram of the
spacetime described above is given in figure 1.  In the section of
spacetime covered by $\lambda^2 x^+ x^- ~ +~ M/\lambda~ >~ 0$,
labeled by I and II, the Killing vector $\xi$,
$$(\xi^+, \xi^-)~~ =~~ \lambda (x^+, - x^-), \eqno(10)$$
is spacelike in region I and timelike in region II.

Since the naked singularity appears at spatial infinity, the ADM
mass is a meaningless concept.  We want to relate $M$ to its Bondi
mass.  There is a standard procedure to evaluate the Bondi energy
of a configuration admitting a Killing vector.  The existence of a
Killing vector implies that the current density defined by
$$j_\mu~~ =~~ {\cal T}_{\mu\nu} \xi^\nu \eqno(11)$$
is conserved.  Let $t_{\mu\nu}$ be a linearization of ${\cal
T}_{\mu\nu}$ about the dilaton vacuum so that to first order $j_\mu
= t_{\mu\nu} \xi^\nu$, and consider a solution of the dilaton field
that is asymptotic to the vacuum with $\phi = \phi^{(0)} +  \delta
\phi$, where $\phi^{(0)} = - \ln (\lambda^2 x^+ x^-)/2$.  The
conserved current density, (11), will take the form
$$\eqalign{j_+~~ &=~~ 2 \lambda \partial_+ \left( e^{-2\phi^{(0)}}
\left[ \delta \phi~ +~ x^+ \partial_+ \delta \phi~ +~ x^-
\partial_- \delta \phi \right] \right) \cr j_-~~ &=~~ 2 \lambda
\partial_- \left( e^{-2\phi^{(0)}} \left[ \delta \phi~ +~ x^+
\partial_+ \delta \phi~ +~ x^- \partial_- \delta \phi \right]
\right). \cr} \eqno(12)$$
The conservation of $j_\mu$ implies the existence of two charges,
$Q^+$ and $Q^-$, which evolve the system in the direction of
increasing $x^+$ and $x^-$.  Integrating, for example, the
conservation equation, $\nabla \cdot j = \partial_+ j_- +
\partial_- j_+ = 0$, over $x^+$ shows that $Q^- = \int^{{\cal
I}_R^+} dx^+ j_+$ is constant in $x^-$.  The current density $j_+$
is a total derivative, so its integral can be measured as a surface
term on $\scrrp$.  Evaluating $\delta \phi$ from (8), one obtains
this conserved charge,
$$Q^-~~ =~~ 2 \lambda \left( e^{-2\phi^{(0)}} \left[ \delta \phi~
+~ x^+ \partial_+ \delta \phi~ +~ x^- \partial_- \delta \phi
\right] \right)_{{\cal I}_R^+}~~ =~~ +~ M~~ >~~ 0. \eqno(13)$$
This is the Bondi mass of the naked singularity, and it is
positive.

The naked singularity described above can be dynamically created by
an incoming pulse of the minimally coupled scalar fields.  Consider
a shock wave${}^7$ of one of the $N$ conformally coupled matter
fields traveling in the $x^-$ direction (at constant $x^+ = x^+_0$)
and set all the other matter fields to zero.  If the shock wave has
magnitude $a$, it is described by the stress tensor
$${1 \over 2}~ \partial_+ f~ \partial_+ f~~ =~~ a \delta(x^+~ -~
x^+_0) \eqno(14)$$
Once again in the gauge $h(x) = 0$, the solution becomes
$$\sigma~~ =~~ \lambda^2 x^+ x^-~ -~ a (x^+~ -~ x^+_0) \Theta (x^+~
-~ x^+_0), \eqno(15)$$
which is just the linear dilaton vacuum when $x^+ < x^+_0$, and the
naked singularity with mass $M = ax^+_0\lambda$ when $x^+ > x^+_0$.
The resulting metric and curvature are identical to those in (8),
with a shift in retarded time, $x^- \rightarrow x^- - a/\lambda^2$.
The Penrose diagram for this spacetime is given in figure 2.  Note
that the naked singularity is formed by the incoming shock wave in
the distant past.  This contrasts with the formation of the black
hole which occurs after the shockwave has traversed a considerable
part of spacetime.

We now want to include quantum effects to the one loop order.  As
is well known, in two dimensions (and only in two dimensions), the
quantum stress energy tensor of evaporation can always be
calculated exactly from the conservation equations and the trace
anomaly.  The latter is given by the well known expression $\langle
T^\mu_\mu \rangle = - 4\sigma \langle T_{+-}\rangle = -~ \alpha R$,
where $\alpha$ is a spin dependent constant which, for bosonic
fields, is equal to $+1/24\pi$.  The conservation equations,
$\nabla_\mu \langle T^\mu_\nu \rangle = 0$, determine the
components of the stress tensor in terms of its trace
$$\eqalign{\langle T_{++}\rangle~~ &=~~ -~ \int~ {{dx^-} \over
\sigma} \partial_+ (\sigma \langle T_{+-}\rangle)~~ +~~ A(x^+) \cr
\langle T_{--}\rangle~~ &=~~ -~ \int~ {{dx^+} \over \sigma}
\partial_- (\sigma \langle T_{+-}\rangle)~~ +~~ B(x^-) \cr}
\eqno(16)$$
where $A(x^+)$ and $B(x^-)$ are boundary condition dependent
functions of $x^+$ and $x^-$ respectively.  For the collapsing $f-$
wave, the natural boundary condition is that the stress tensor
vanishes in the linear dilaton vacuum.  This fixes both $A(x^+)$
and $B(x^-)$ giving, for the components of $\langle T_{\mu\nu}
\rangle $ the final expressions
$$\eqalign{\langle T_{++}\rangle~~ &=~~ +~ {{\alpha \lambda^4 (x^-
- a/\lambda^2)^2} \over {2 \sigma^2}}~ -~ {\alpha \over {2
{x^+}^2}} \cr \langle T_{--}\rangle~~ &=~~ +~ {{\alpha \lambda^4
{x^+}^2} \over {2 \sigma^2}}~ -~ {\alpha \over {2 {x^-}^2}} \cr
\langle T_{+-}\rangle~~ &=~~ +~ {{\alpha \lambda^2 a x^+_0} \over
{\sigma^2}}. \cr} \eqno(17)$$
It is most convenient to analyze the above expressions in the
coordinate system in which the metric is asymptotically flat.  In
region I of figure 2, define the new coordinates $\sigma^\pm = t
\pm x$ by
$$\eqalign{x^+~~ &=~~ {1 \over \lambda}~ e^{\lambda \sigma^+} \cr
x^-~~ &=~~ {1 \over \lambda}~ e^{\lambda \sigma^-}~ +~ {a \over
{\lambda^2}}. \cr} \eqno(18)$$
Thus, $\sigma^- \rightarrow - \infty$ corresponds to the light-like
surface at $x^- = a/\lambda^2$ shown in  figure 2. This
transformation preserves the conformal gauge, and gives for the new
metric in region I
$$ds^2~~ =~~ {{dt^2~ -~ dx^2} \over {1+ax^+_0 e^{-2\lambda
t}}}\eqno(19)$$
Transforming the expressions above for $\langle T_{\mu\nu} \rangle$
to the new coordinate system, we find on $\scrrp$ (as $\sigma^+
\rightarrow \infty$)
$$\eqalign{\langle T^{(\sigma)}_{++}\rangle~~ & \rightarrow~~
0,~~~~~~~~~~ \langle T^{(\sigma)}_{+-} \rangle~~ \rightarrow~~ 0
\cr \langle T^{(\sigma)}_{--}\rangle~~ &=~~  {{\alpha \lambda^2}
\over 2}~ \left[ 1~~ -~~ {1 \over {(1~ +~ (a/\lambda) e^{-\lambda
\sigma^-})^2}} \right]. \cr} \eqno(20)$$
Now, $\langle T^{(\sigma)}_{--} \rangle$ represents the outgoing
flux at $\scrrp$.  It approaches a maximum of $\alpha \lambda^2 /2$
in the far past of $\scrrp$ at $x^- = a/\lambda^2$, and decreases
smoothly to zero in the far future of $\scrrp$ as $x^- \rightarrow
\infty$.  All components of the tensor vanish on $\scrlp$.

This is the Hawking radiation from the naked singularity.  At $x^-
= a/\lambda^2$ on $\scrrp$, it is independent of the mass of the
singularity just as the radiation from two dimensional black holes
approaches a constant independent of $M$ near the horizon.  The
total energy lost by the collapsing $f-$ wave at a fixed retarded
time, $\sigma^-$, is the integrated flux along $\scrrp$ from
$\sigma^- \rightarrow - \infty$ to $\sigma^-$.  This integrated
flux is infinite because the flux approaches a steady state at
early retarded times.  Of course, the singularity cannot radiate
away more energy than it possesses.  The result is nonsense, a
consequence of having neglected the back reaction of the radiation
on the spacetime geometry.

It is safe to believe, however, that the radiation process is
cataclysmic, and that the $f-$ wave radiates away most of its
energy at a point very close to $(x^+_0, a/\lambda^2)$.  To justify
more quantitatively this statement, consider the value of the
dilaton field at this point, keeping in mind that $e^\phi$ is the
loop expansion parameter for dilaton gravity.  One finds
$$e^\phi~~ =~~ {1 \over {\sqrt{ax^+_0}}} \eqno(21)$$
which is indeed small for very energetic incoming $f-$ shock waves
(large $a$), making the one loop approximation credible in this
case.  The situation here contrasts starkly with the black hole. In
the latter case, the flux approaches zero at early retarded times
approaching a constant as the horizon is approached on $\scrrp$.
It was shown in ref(7) that the dilaton coupling is not small at
the point at which the $f-$ wave is expected to radiate itself
away, implying that the one-loop calculation breaks down before the
$f-$ wave fully disappears.  Likewise, for small mass naked
singularities, the loop expansion parameter is large and the
results of a one-loop approximation are not credible.  Moreover,
given a fixed, small mass naked singularity, proliferating the
number of matter fields cannot remedy the situation because the
presence of $N$ matter fields serves only to multiply the Hawking
radiation by a factor of $N$.  The naked singularity therefore
evaporates even more catastrophically.

One might attempt to include quantum effects in our earlier
computation of the configuration's Bondi energy.  To do so, one
must add the quantum stress tensors derived above to the
expressions we had for $t_{\mu\nu}$.  The quantum corrected
conserved current density is thus the sum $j_\mu~~ =~~
j^{(old)}_\mu~~ +~~ j^Q_\mu$, where $j^Q_\mu$ is due to the Hawking
radiation, and $j^{(old)}_\mu$ is given in (12) after the
appropriate shift in retarded time.  Linearizing as before, the
charge density is once again found to be a total derivative, and
the mass of the naked singularity appears as a surface term on
$\scrrp$,
$$\eqalign{M(x^-)~~ &=~~ 2 \lambda \left( e^{-2\phi^{(0)}} \left[
\delta \phi~ +~ x^+ \partial_+ \delta \phi~ +~ (x^- - {a \over
{\lambda^2}}) \partial_- \delta \phi \right ] \right. \cr & \left
. ~~~~~~~~~~~~~ +~ \alpha \left[ x^+ \partial_+ \delta \phi~ +~
(x^- - {a \over {\lambda^2}}) \partial_- \delta \phi \right ]
\right )_{{\cal I}_R^+}~~ =~~ ax^+_0 \lambda. \cr} \eqno(22)$$
It is, however, independent of $x^-$ because the term proportional
to $\alpha$ vanishes as a consequence of the boundary conditions
which imply that there are no geometric invariants on $\scrrp$.
This would seem to lead to the absurd conclusion that the quantum
corrected Bondi energy is constant despite the evaporation, but it
is once again a consequence of having neglected the back reaction
on the spacetime geometry.

The picture that seems to emerge from the above is the following.
If the full quantum theory permits the formation of a naked
singularity, it will evaporate catastrophically (``explode'') due
to its Hawking radiation.  For large mass naked singularities, the
one loop anomaly calculation is believable, as the loop expansion
parameter at the explosion point is small.  In a full quantum
treatment however, the $f-$ wave may not instantly give up all its
energy to the Hawking radiation as one is led to believe from the
above. A large mass naked singularity will rapidly, but smoothly,
give up its energy, diminishing in size in a short time.  Because
of the radiation, we expect that the mass of the singularity will
be a function, $M(\sigma^-)$, of the retarded time $\sigma^-$ (or
$x^-$), that is, $M(\sigma^-)$ should approximate zero very shortly
in the retarded future of $a/\lambda^2$ on $\scrrp$.  As
$M(\sigma^-)$  becomes small, the theory becomes strongly coupled
and the one loop approximation breaks down.  Without a full quantum
treatment, it is therefore premature to predict its final fate.

One result that is expected to prevail in four dimensions is that
if a naked singularity forms it will evaporate catastrophically.
As the evaporation starts in the far past, at the approach to the
naked singularity, quantum mechanical effects may actually prevent
the latter from forming.  Thus one can imagine that quantum cosmic
censorship occurs either by preventing the formation of a naked
singularity or causing it to explode immediately after formation.
In the second alternative the naked singularity would be an
``event'' in spacetime${}^{13}$.
\vskip 0.2in
\noindent{\bf Acknowledgement}

\noindent This work was supported in part by  NATO under contract
number CRG 920096. L.W. acknowledges the partial support of the U.
S. Department of Energy under contract number DOE-FG02-84ER40153.
\vskip 0.2in
\noindent{\bf Figure Captions:}

\noindent{\bf 1.} Figure 1. Penrose diagram of the naked
singularity represented by (8)

\noindent{\bf 2.} Figure 2. Penrose diagram of the naked
singularity formed by an incoming $f-$ shock wave.
\vskip 0.2in

\noindent{\bf References:}

{\item{1.}}S. W. Hawking, Comm. Math. Phys {\bf 43} (1975) 199.

{\item{2.}}J. B. Hartle and S. W. Hawking, Phys. Rev. {\bf D13}
(1976) 2188; B. S. DeWitt, Phys. Rep. {\bf 19C} 295 (1975); W.
Israel, Phys. Letts. {\bf 57A} (1976) 107; W.G. Unruh, Phys. Rev.
{\bf D14} (1976) 840; S. M. Christensen and S. A. Fulling, Phys.
Rev {\bf D15} (1977) 2088; James Bardeen, Phys. Rev. Lett. {\bf 46}
(1981) 382; D.N. Page, Phys. Rev. {\bf D25} (1982) 1499; K. W.
Howard and P. Candelas, Phys. Rev. Lett. {\bf 53} 403 (1984);
Cenalo Vaz, Phys. Rev. {\bf D39} 1988 1776; Fernando A. Barrios and
Cenalo Vaz; Phys. Rev. {\bf{D40}} (1989) 1340.

{\item{3.}}S. W. Hawking, Phys. Rev. {\bf D14} (1976) 2460.

{\item{4.}}R. Jackiw in {\it Quantum Theory of Gravity}, ed. S. M.
Christensen (Hilger, Bristol) 1984; A. H. Chamseddine, Phys. Letts
{\bf B256} (1991) 379; R. B. Mann, A Shiekh and L. Tanasov, Nucl.
Phys. {\bf B341} (1990); R. B. Mann, Found. of Phys. Letts. {\bf 4}
(1991) 425; D. Garfinkle, G Horrowitz and A. Strominger, Phys. Rev.
{\bf D43} (1991) 3140; G. Horrowitz and A. Strominger, Nucl. Phys.
{\bf B360} (1991) 197; S. B. Giddings and A. Strominger, Phys. Rev.
Letts. {\bf 67} (1991) 2930.

{\item{5.}}G. Mandal, M. M. Sengupta and S. R. Wadia, Mod. Phys.
Letts. {\bf A6} (1991) 1685.

{\item{6.}}E. Witten, Phys. Rev. {\bf D44} 314 (1991).

{\item{7.}}C. Callan, S. Giddings, J. Harvey and A. Strominger,
Phys. Rev {\bf D45} (1992) 1005.

{\item{8.}}A. Bilal and C. Callan, Princeton University preprint
PUPT-1320; S. P. de Alwis, Phys. Letts. {\bf 289B} (1992) 278; {\it
ibid} Phys. Rev. {\bf D46} (1992) 5429; J. G. Russo, L Susskind and
L. Thorlacius, Phys. Rev. {\bf D46} (1992) 3444; Phys. Rev. {\bf
D47} (1993) 533; A. Strominger, Phys. Rev. {\bf D46} (1992) 4396;
K. Schoutens, H. Verlinde and E. Verlinde, Princeton University
preprint PUPT-1395.

{\item{9.}}S. Weinberg, {\it Gravitation and Cosmology} (John Wiley
\& Sons, Inc., New York) (1972).

{\item{10.}}C. H. Brans and R. H. Dicke, Phys. Rev. {\bf 124}
(1961) 925; R. H. Dicke, Phys. Rev. {\bf 125} (1962) 2163.

{\item{11.}}C. Nappi in {\it Topics in Quantum Gravity and Beyond},
Proceedings of the Conference in Honor of Louis Witten, eds.
Freydoon Mansouri and Joseph J. Scanio (World Scientific, New
Jersey) 1993; Jos\'e P. S. Lemos and Paulo M. S\'a, Instituto
Superior T\'ecnico preprint DF/IST-9.93, August 1993.

{\item{12.}}H. Bondi, M. G. J. van der Burg and A. W. K. Metzner,
Proc. Roy. Soc. {\bf A269} (1962) 21.

{\item{13.}}E. Witten in {\it Topics in Quantum Gravity and
Beyond}, Proceedings of the Conference in Honor of Louis Witten,
eds. Freydoon Mansouri and Joseph J. Scanio (World Scientific, New
Jersey) 1993.
\end